\documentstyle[12pt,aasms4]{article}

\received{}
\accepted{}
\journalid{}{}
\articleid{}{}

\slugcomment{Accepted by ApJ Letters}

\lefthead{Martin et al.}
\righthead{The Lithium Substellar Boundary in the Pleiades}

\begin{document}

\title{A New Pleiades Member at the Lithium Substellar Boundary}

\author{E.L. Mart\'\i n and G. Basri}
\affil{Astronomy Department, University of California,
    Berkeley, CA 94720}

\and

\author{J.E. Gallegos, R. Rebolo, M.R. Zapatero-Osorio, V.J.S. Bejar}
\affil{Instituto de Astrof\'\i sica de Canarias, 38200 La Laguna, Spain}


\centerline{e-mail addresses: ege@popsicle.berkeley.edu, basri@soleil.berkeley.edu} 
\centerline{jgallego@ll.iac.es,rrl@ll.iac.es,mosorio@ll.iac.es,vbejar@ll.iac.es}

\begin{abstract}
We present the discovery of an object in the Pleiades open cluster, named  
Teide~2, with optical and infrared photometry which place it on the cluster 
sequence slightly below the expected substellar mass limit. We have obtained 
low- and high-resolution spectra that allow us to determine its 
spectral type (M6), radial velocity and rotational broadening; and to detect    
H$_\alpha$ in emission and Li\,{\sc i} 670.8~nm in absorption. 
All the observed properties strongly support the membership of 
Teide~2 into the Pleiades. This object has an important role 
in  defining the reappearance of lithium below the substellar limit in the Pleiades. The age of the Pleiades very low-mass members  
based on their luminosities and absence or presence of 
lithium is constrained to be in the range 100--120~Myr.

\end{abstract}


\section{Introduction}

The Pleiades open cluster has become a favorite location for  
 brown dwarf (BD) searches. This privilege is due to its proximity, 
young age, low extinction and compactness (the same properties 
that make it easily recognized to the naked eye). 
Such advantages were 
recognized over ten years ago, when the first CCD surveys of the cluster 
began (Jameson \& Skillen \cite{jameson89}; 
Stauffer et al. \cite{stauffer89}). Since then, 
there has been many efforts to reach fainter magnitudes and cover 
more area (see Hambly \cite{hambly98} for a review), which have revealed 
many BD candidates. 

A breakthrough in our understanding of BDs has come from the successful 
application of the lithium test 
(Rebolo, Mart\'\i n \& Magazz\`u \cite{rebolo92}; Magazz\`u, Mart\'\i n \& 
Rebolo \cite{magaz93}), 
which is an efficient observational method  
for distinguishing young BDs from very low-mass (VLM) stars. 
The first object at the substellar limit with a Li detection was 
PPl 15 (Basri, Marcy \& Graham \cite{basri96}; hereafter BMG) 
and soon after Li was detected in the BDs  
Teide 1 and Calar 3 (Rebolo et al. 1996). 
The detection of 
lithium in PPl~15 and the non-detection in HHJ~3 (Marcy et al. 1994) 
was used by BMG to 
infer an age of $\sim$115~Myr for the Pleiades cluster, 
which is substantially older than the 
canonical cluster age of 75~Myr inferred from standard calculations of the 
upper main-sequence turnoff. They argued that the turnoff age 
increases when core convective overshoot is included in the evolutionary  
calculations, an effect that could account for the difference between dating 
the cluster with high-mass stars and with VLM objects. 

PPl~15 was shown to lie on the binary infrared cluster sequence 
(Zapatero-Osorio, Mart\'\i n \& Rebolo 
\cite{osorio97a}), and recently it has been found to be a spectroscopic double-lined BD 
binary (Basri \& Mart\'\i n \cite{basri98}). Thus, 
the luminosity used by BMG was overestimated because 
they assumed that the object was single. This is relevant for 
the discussion of the cluster age, because the luminosity and lithium 
of PPl~15 implied that it had to be younger than 125~Myr. Since the 
luminosity of the primary in the PPl~15 binary is lower, the age could be 
older. In this paper we present the discovery of a new object in the Pleiades, 
which has a luminosity similar to the PPl~15 system. This object is located 
on the cluster sequence for single objects, and thus its luminosity is 
not expected to be affected by a close companion. We have detected 
lithium in it, which confirms its cluster membership, and makes it 
an important object for defining the  
lithium substellar boundary and constraining the cluster age.

\section{Observations}

The object presented in this paper was discovered in a CCD survey of the 
Pleiades carried out using the 0.82~m IAC80 telescope at the Observatorio 
del Teide in November and December of 1995, and January 1996. 
We used a Thomson 1048$^2$ CCD with a field of view of 
54.5 arcmin$^2$ and $R,I$ filters. Two exposure of 1800s 
were obtained for each filter. 
We observed 13 different fields, forming a mosaic   
of 590~arcmin$^2$. 
The images were bias subtracted, flat fielded, corrected for bad 
columns, and searched for pointlike sources using DAOPHOT routines 
within IRAF\footnote {IRAF is distributed
by National Optical Astronomy Observatory, which is operated by the Association
of Universities for Research in Astronomy, Inc., under contract with the
National Science Foundation.}. The FWHM of the stars in our frames were 
in the range 1".5 to 2".5. The instrumental magnitudes were calibrated 
with a secondary standard field within the Pleiades that we had observed in 
previous runs (Zapatero-Osorio, Rebolo \& Mart\'\i n \cite{osorio97b}). 
Due to variable seeing conditions, we did not reach 
the same sensitivity in all the fields. We estimate that our completeness 
limits are in the range $I$=18.5--19.0 mag. 

Near-infrared (NIR) follow-up observations of the BD candidates found 
in the CCD survey were 
carried out on 1996 Nov 2 and 1997 Sep 21--23 at the 
Calar Alto 2.2~m telescope with MAGIC, and the Lick 1~m telescope with 
LIRCII, respectively. The standard $J$ and $K'$ filter set was 
used. Total exposure times ranged from 60 to 120~s in each filter. 
We used the IRAF DAOPHOT package to reduce the frames and extract the 
photometry. Instrumental aperture magnitudes were corrected for 
atmospheric extinction and transformed into the CIT system using 
observations of the stars in the field of the BD Calar~3 as standards 
 (Zapatero-Osorio et al. \cite{osorio97a}). 
Only one BD candidate had optical 
and infrared photometry consistent with being a Pleiades member. 
We named it Teide~Pleiades~2 (from now on abreviated simply to Teide~2)
because it resulted from the extension of the Pleiades survey at the Teide Observatory.  
A preliminary report of this discovery was made by Gallegos et al.  
 (\cite{gallego98}). 
The coordinates and $R,I,J,K$ photometry are listed 
 in Table~\ref{tab1}  and the finding chart is shown in Figure~\ref{fig1}. 
Teide~2 is included in the recent CFHT survey 
(Bouvier et al.  \cite{bouvier98}), where it has a magnitude of 
$I$=18.02$\pm$0.05 which is 
0.2 mag brighter than ours. We have compared directly the CFHT and IAC80 
CCD images 
and find that Teide~2 has changed its brightness with respect to 
field reference stars. For the J-band, the  
magnitudes obtained in Calar Alto and Lick 
differ by 0.12 mag, which is somewhat larger than the expected 
error bars. Teide~2 is probably variable in $I$, and perhaps 
also in $J$. This suggestion needs to be confirmed with 
photometric monitoring, which might measure the rotation 
period if the variability is due to photospheric spots.   

Low-resolution spectroscopy of Teide~2 was obtained on 1996 Dec 8 and 9 
using the  ISIS double-arm spectrograph at the WHT 
(we only used the red arm) with the grating R158R and a TEK 
1024$^2$ CCD detector; and the LRIS spectrograph 
with the 1200~line~mm$^{-1}$ grating and a TEK 
2048$^2$ CCD detector at the Keck~II telescope. 
The spectral resolution (FWHM) and wavelength coverage provided 
by each instrumental 
setup were 5.8~\AA ~ and 650--915~nm , and 1.9~\AA ~ and 654--775~nm , 
respectively. Reduction and calibration procedures were the same 
as described in Zapatero-Osorio et al. (\cite{osorio97c}). 
High-resolution spectroscopic observations were obtained on 
1997 Dec 3 and 7, at the Keck~I telescope equipped with 
the HIRES echelle spectrometer. The instrumental configuration and 
data reduction have been discussed by Marcy et al. (1994) and BMG. 
 The resolving power obtained was R=31000. 
We took two spectra of Teide~2 with exposure times of 3600~s and 4000~s. 
However, the sky was cloudy when the first spectrum was taken and 
we obtained a factor of 2 less counts than in the second spectrum. 
We also obtained one 2700~s exposure of PPl~1, a BD candidate discovered 
by Stauffer et al. (1989) which has a similar brightness to Teide~2.  

\section{Results and discussion}

A color-magnitude diagram displaying the position of Teide~2 and PPl~1 
with respect to other members of the Pleiades is 
presented in Figure~\ref{fig2} together with a theoretical 120~Myr isochrone 
(dashed line), and the main-sequence 
of field M-dwarfs (solid line) shifted to the Pleiades distance (127~pc) and 
mean reddening (A$_{\rm V}$=0.12). Teide~2 falls close to the theoretical isochrone, 
whereas PPl~15 clearly lies above it, suggesting that it could be a binary 
(Zapatero-Osorio et al. 1997a). 
Basri \& Mart\'\i n (1998) confirmed that PPl~15 is indeed a double-lined 
spectroscopic binary with a mass ratio close to unity, implying that both 
components have nearly identical luminosities. Thus, we have indicated  
in Figure~\ref{fig2} that the I magnitude of each component in the PPl~15 
system is $\sim$0.75 mag. fainter, while the $I-K$ color remains 
about the same. 
PPl~15 now falls right on the 120~Myr isochrone for a mass of 0.06~M$_\odot$,  
well below the substellar boundary. 
The location of Teide~2 in Figure~\ref{fig2} indicates a mass of 
0.072~M$_\odot$ which is just below the expected substellar limit 
(0.075~M$_\odot$, cf. Baraffe et al. 1998). 
The position of PPl~1 above the isochrone might be 
due to measurement errors, photometric variability, binarity or younger 
age. We cannot presently distinguish between these possibilities.    

Radial velocities for Teide~2 and PPl~1 were obtained by 
cross-correlating three HIRES echelle orders covering 
the ranges 786.4--797.5~nm, 804.2--815.6~nm and 822.9--834.6~nm with 
the M6V template Gl~406 ($v_{\rm rad}$=19.2$\pm$0.1 km s$^{-1}$; 
X. Delfosse, private communication) observed using the same 
instrumental configuration. The radial velocities obtained are given 
in Table~\ref{tab1}. The error bars listed  
come from the dispersion among the different echelle orders, 
and correspond to about 0.4 pixels. No significant radial velocity 
variation was detected between the two spectra of Teide~2 taken 
on Dec~3 and Dec~7. Stauffer et al. (\cite{stauffer94}) obtained a 
radial velocity for PPl~1 of 14.3$\pm$10 km s$^{-1}$, which is consistent 
with our measurement. The radial velocity of Teide~2 is consistent  
with known cluster members, while that of PPl~1 is a little 
on the high side. The mean velocity of nearby field 
dwarfs in the line of sight toward the Pleiades is also similar to the cluster 
members (Stauffer et al. \cite{stauffer94}), hence radial velocity alone is not very reliable as a membership criterion.  

There are more spectroscopic indicators that allow us 
to further investigate 
the membership of our objects in the Pleiades. One is the strength of 
H$\alpha$ in emission, which is typically greater than 3~\AA ~ in proper 
motion M4--M6 cluster members (Hodgkin, Jameson \& Steele \cite{hodgkin95}).  
For Teide~2 we measured H$\alpha$ equivalent widths of 9.5$\pm 1.0$, 
8.5$\pm 1.0$ and 4.7$\pm 0.4$~\AA ~ 
in our ISIS, LRIS and HIRES spectra, respectively. For PPl~1 we measured 
2.3$\pm 1.5$\AA ~ in our HIRES spectrum.  Stauffer et al. (\cite{stauffer94}) 
and Mart\'\i n, Rebolo \& Zapatero Osorio (\cite{martin96}) 
published H$\alpha$ measurement for PPl~1 
of 4.8~\AA ~ and 6.5~\AA , respectively. 
The H$\alpha$ emission of both Teide~2 and PPl~1 has 
variable strength, but on the average they are comparable to those of late M-type proper motion Pleiades members. 
We measured the rotational 
broadening of Teide~2 and PPl~1 in our HIRES spectra. 
We artificially broadened Gl~406 (vsin{\it i}$<$2.9~km~s$^{-1}$; Delfosse et al. \cite{delfosse98}) with different 
values of vsin{\it i} until we found the best match to the observed 
broadening of our Pleiades objects. The results are 
given in Table~\ref{tab1}. 
Field M4--M6 dwarfs usually have  vsin{\it i}$<$10 
(Delfosse et al. \cite{delfosse98}). 
Thus, the vsin{\it i} of Teide~2 and PPl~1 
also indicate that they are young. A third clue to membership comes  
from the general features seen in the low-resolution spectrum of Teide~2, 
which shows prominent TiO and VO molecular absorption bands and strong 
atomic lines of K\,{\sc i} (766.5 and 769.9~nm) and Na\,{\sc i} 
(818.3 and 819.5~nm) characteristic of a cool dwarf. 
The spectral type of Teide~2 is M6$\pm$0.5 according to the 
pseudocontinuum indices defined by (Mart\'\i n et al. \cite{martin96}). 
Teide~2 nicely fits the sequence of Pleiades members in the 
$I$-magnitude versus spectral type diagram of Mart\'\i n et al. 
(\cite{martin96}). Summarizing, the radial velocities, H$\alpha$ emission, 
vsin{\it i}, spectral type and photometric properties of Teide~2 and PPl~1 
support their Pleiades membership with a high level of confidence.  

In the top half of Figure~\ref{fig3} we compare the spectra of PPl~1, Teide~2 
and PPl~15. The spectrum of PPl~1 is noisier than those of the other two 
objects because it was taken with the 2.5~m Isaac Newton telescope 
(Mart\'\i n et al. \cite{martin96}). Teide~2 and PPl~1 have similar spectral type within the error bars, 
and PPl~15 is somewhat cooler but not more than one spectral subclass. 
Teide~2 confirms that   
the dividing line between stars and BDs in the Pleiades is at spectral type 
$\sim$M6 (Mart\'\i n et al. \cite{martin96}). 
This result strengthens 
the recent identification of objects M6--M8 in several star-forming regions 
(Orion, Taurus, $\rho$~Oph) as likely brown dwarfs. Brice\~no 
et al. (\cite{briceno98}) and Luhman et al. (\cite{luhman98}) have shown that the Na\,{\sc i} lines of some BD candidates in Taurus are much weaker than those of dwarfs, indicating 
that the very young BDs have low gravities. We note that the 
equivalent width of the Na\,{\sc i} (818.3 and 819.5~nm) measured in our 
WHT spectrum of Teide~2 is 
6.0$\pm$0.3~\AA . This line is stronger than those of the 
Taurus BDs of similar spectral type as expected because the Pleiades is 
much older, but slightly weaker than those of old 
M6 field dwarfs.   

In the bottom half of Figure~\ref{fig3} we compare the HIRES spectra of 
HHJ~3 and Teide~2. Marcy et al. (1994) placed an upper limit to the 
Li\,{\sc i} equivalent width of HHJ~3 of 0.19~\AA . Our spectrum of Teide~2 
has lower S/N ratio than their spectrum of HHJ~3, 
because our target is fainter and we had worse weather conditions. 
However, the Li\,{\sc i} feature in Teide~2 is clearly much stronger 
than in HHJ~3 and we were able to detect it in our spectra. We measured 
an equivalent width of 0.77$\pm$0.15~\AA ~ with respect to the local pseudocontinuum. Our spectrum of PPl~1 has a factor of about 2 lower S/N ratio than that of Teide~2. 
We degraded the resolution by performing a rebinning of 6 pixels 
and found a feature 
at the position of the Li\,{\sc i} line. However, we are not completely confident 
of a lithium detection in PPl~1, and hence we place   
a conservative upper limit to its equivalent width. 
We have compared the spectrum of Teide~2 with that of 
PPl~15 (BMG) and we conclude that their 
Li\,{\sc i} strengths are similar. The conversion from the observed 
lines to photospheric lithium abundance needs the use of synthetic spectra. 
The main problem lies with our incomplete knowledge of the transitions 
of the TiO molecule, which is the main absorber in this spectral region. 
New calculations by Pavlenko (\cite{pav97}) suggest that the lithium 
abundances of Teide~2 and PPl~15 could be close to the interstellar value, 
implying that these objects have suffered little depletion. The estimate 
made by BMG that PPl~15 had preserved $\sim$1\% 
of its initial lithium is a conservative lower limit. Furthermore, 
their measure of the Li\,{\sc i} equivalent width is compromised by its binary nature.

The VLM members of the Pleiades ($\le$0.1~M$_\odot$) are 
non-degenerate, fully convective objects that 
derive most of their luminosity from gravitational contraction. 
Hence, their evolution is relatively simple and 
can even be computed analytically (Bildsten et al. \cite{bildsten97}). 
The faintest 
Pleiades member for which lithium is undetected gives a lower limit 
to the age. This is currently HHJ~3, and  
BMG obtained an age older than about 110~Myrs for it. 
We have checked with new theoretical models (Baraffe et al. \cite{baraffe98}; 
Bildsten et al. 1997; Burrows 1997 (private communication); D'Antona \& Mazzitelli \cite{dantona98}) that the lower age limit set by  
HHJ~3 is in the range 100--120~Myr taking into account the observational 
error bars. This result is not sensitive to 
the choice of theoretical models. 
On the other hand, the brightest BDs for 
which lithium is detected provide an upper limit to the cluster age. 
We have converted the $I,K$ photometry and spectral type of Teide~2 
to bolometric luminosity by employing relationships derived for cool field dwarfs (Tinney et al. \cite{tinney93}). Since Teide~2 shows lithium, 
its luminosity implies that it has to be younger than 120~Myr according to 
all the available models. We have also compared directly the absolute 
$I,J,K$magnitudes of Teide~2 with the predictions of Baraffe et al. (1998) 
and find similar results. 
Teide~2 and HHJ~3 can only be coeval if both of them have ages in the  
narrow range 100--120~Myr. With only a few objects around the cluster 
substellar limit we cannot rule out 
the possibility of an age spread, but their location close to the 120~Myr 
isochrone in Figure~\ref{fig2} indicates that they are essentially coeval.  
All the recent models also agree that the mass of Teide~2 is just below the 
substellar limit. Taking into account the observational error bars and 
the different models, we estimate a mass of 0.070$\pm$0.005~M$_\odot$.  
Teide~2 is most likely a single BD, or if it is a 
binary the secondary does not contribute much to the optical-NIR fluxes. 
Hence, Teide~2 can be considered a benchmark object defining  
the borderline between BDs and stars in the Pleiades. 

\acknowledgments

{\it Acknowledgments}: 
This research is based on data collected at the 
W.~M. Keck Observatory, which is operated jointly by the University of 
California and CALTECH; the WHT telescope operated by the Isaac 
Newton Group funded by PPARC at the  Spanish 
Observatorio del Roque de los Muchachos of the Instituto de 
Astrof\'\i sica de Canarias; the 2.2~m telescope at the 
German-Spanish Observatorio de Calar Alto; the Nickel 1~m telescope 
at Lick Observatory run by the University of California;  
and the IAC80 telescope at  
the Observatorio del Teide (Tenerife, Spain). 
We thank Drs. I. Baraffe, A. Burrows, 
X. Delfosse and F. D'Antona for sending results in advance of publication. 
EM acknowledges the support from the F.P.I. program 
of the Spanish Ministry of Education and Culture. 
GB acknowledges the support of NSF through grant AST96-18439.

\clearpage

\figcaption[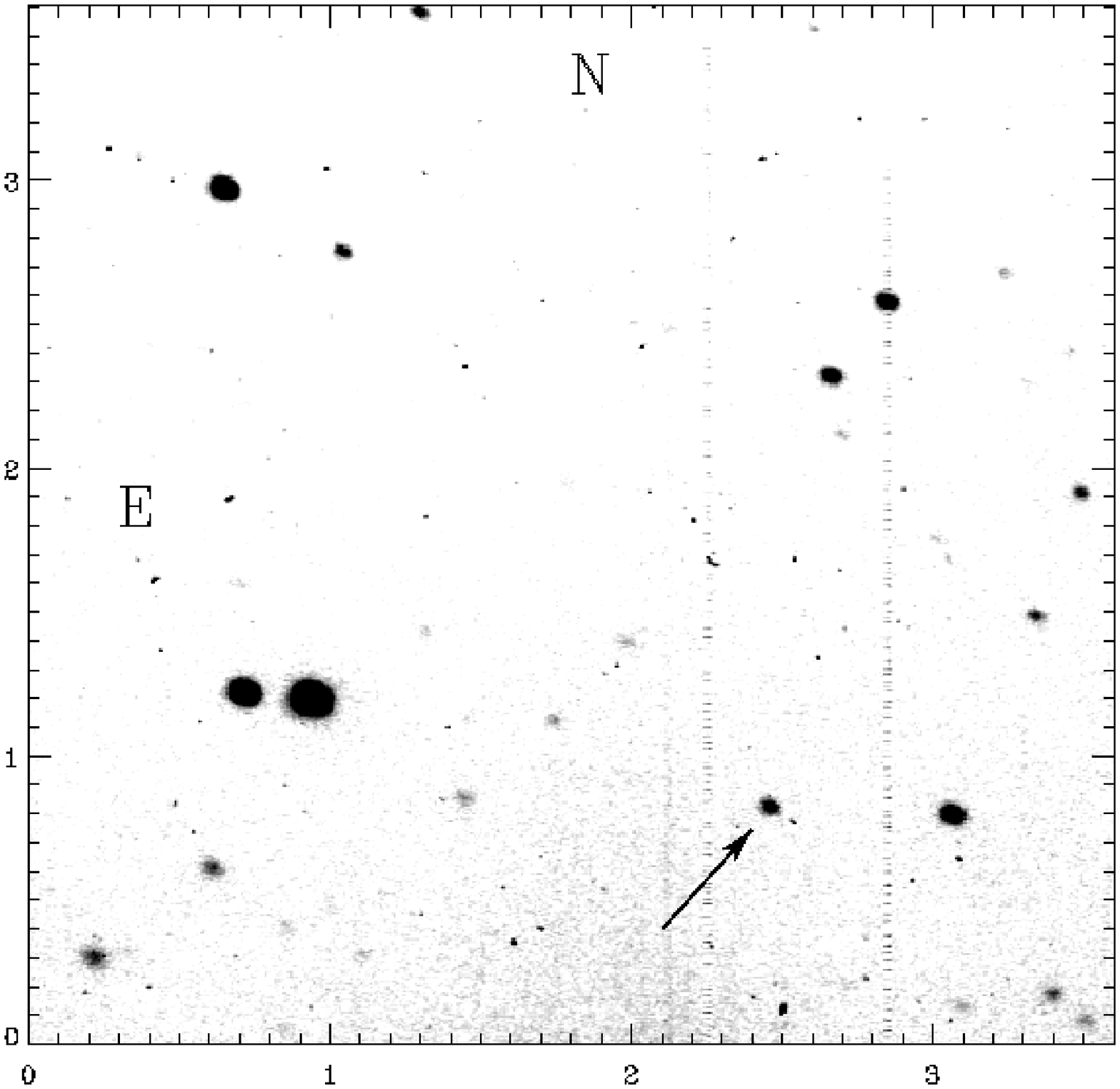]{\label{fig1} I-band finding chart for Teide~2. 
Each side is 3.6 arcmin long.}

\figcaption[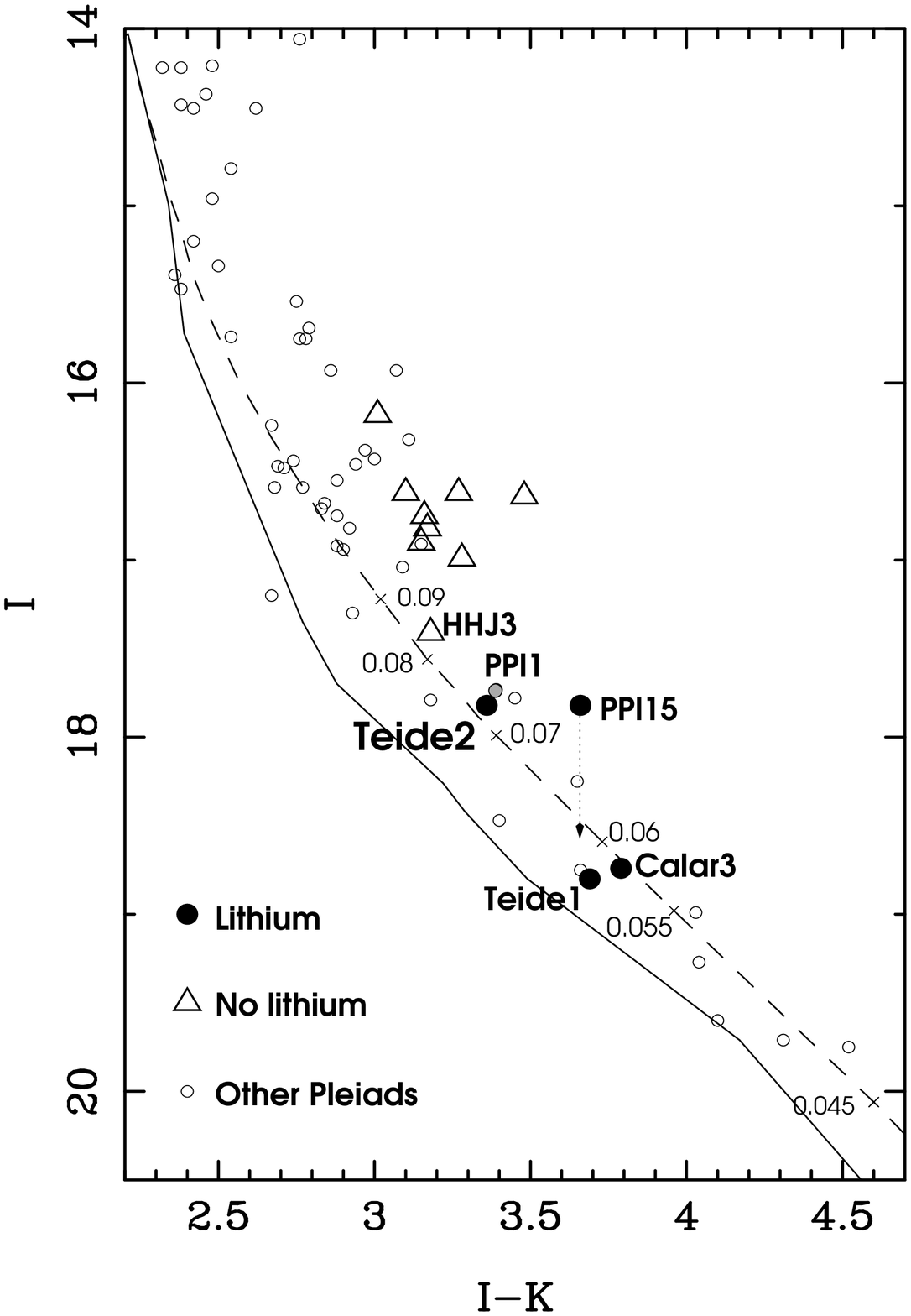]{\label{fig2} The $I$ vs ($I-K$) sequence in 
the Pleiades. Filled circles represent lithium-confirmed BDs 
(BMG, Rebolo et al. 1996, this work).  The arrow 
from PPl~15 indicates that it is a binary with nearly identical components 
(Basri \& Mart\'\i n 1998). The empty triangles are Pleiades proper 
motion members from Hambly et al. (1993) with 
Li\,{\sc i} non-detections (Marcy et al. 1994; Oppenheimer 
et al. 1997). The empty circles are Pleiades objects 
(Zapatero-Osorio et al. 1997c and references therein)
that do not have lithium observations. 
The dashed line corresponds to the 120~Myr isochrone 
(Baraffe et al. 1998) and is labelled with masses in solar units. 
The solid line is the sequence of field VLM dwarfs 
shifted to the Pleiades distance.}

\figcaption[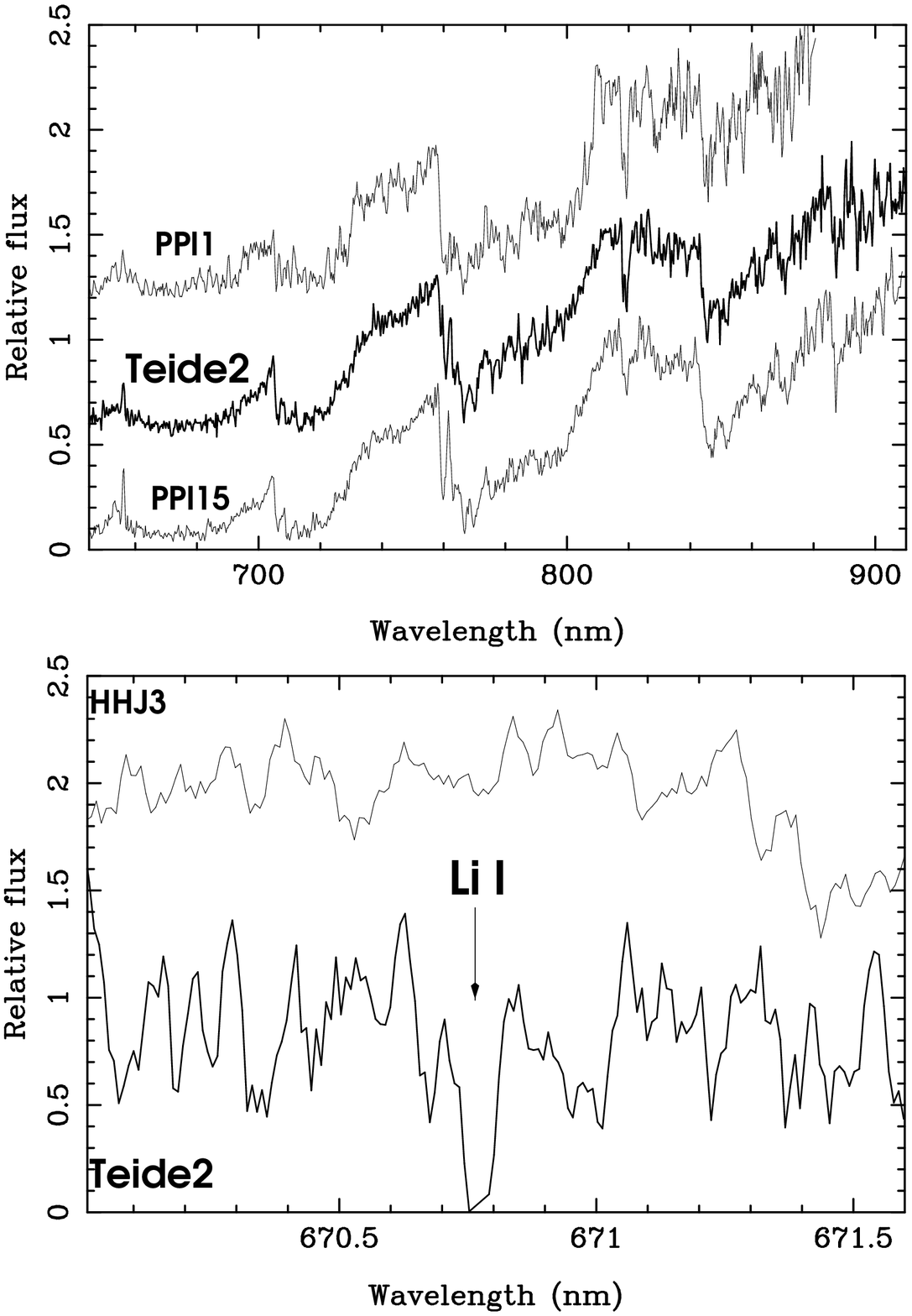]{\label{fig3} Top panel: 
Low-resolution spectra of Teide~2 together with PPl~1 and PPl~15. 
Bottom panel: High-resolution spectra of HHJ~3 and Teide~2 
around the Li\,{\sc i} resonance line region. The data of both objects 
have been boxcar smoothed by 5 pixels.}

\clearpage

\begin{deluxetable}{lcc}
\footnotesize
\tablecaption{\label{tab1} Data for our program objects}
\tablewidth{0pt}
\tablehead{
\colhead{}  & 
\colhead{PPl~1}  &
\colhead{Teide Pleiades 2} \nl   
                            }
\startdata
RA$_{\rm J2000}$ & & 3h 52m 06.7s $\pm0.1s$ \nl
DEC$_{\rm J2000}$ & & 24$^{\circ}$ 16' 01'' $\pm1''$ \nl
$R_{C}$ & & 20.05$\pm 0.10$ \nl
$I_{C}$ & 17.73 & 17.82$\pm 0.05$ \nl
$J_{CIT}$ & 15.42 & 15.45$\pm 0.05$ \nl
$K_{CIT}$ & 14.34 & 14.46$\pm 0.07$ \nl
SpT & M6.5$\pm 1.0$ & M6$\pm 0.5$ \nl
EW (Li\,{\sc i}) (\AA) & $\le$1.8 & 0.77$\pm 0.15$ \nl 
EW (H$\alpha$) (\AA) &  6.5--2.3 & 9.5--4.7 \nl
$v_{\rm rad}$ & 15.4$\pm 1.6$ & 11.2$\pm 1.0$ \nl
$v$sin{\it i} & 18.5$\pm 1.5$ & 13.0$\pm 1.0$ \nl
log~$L/L_{\odot}$ & $-2.85\pm0.03$ & $-2.90\pm0.03$ \nl
\enddata
\tablenotetext{}{For PPl~1 the coordinates are given in 
Stauffer et al. (1989), and 
the magnitudes are the mean between those provided by 
the same authors and Zapatero-Osorio et al. (1997a)}
\end{deluxetable}

\clearpage

\plotone{teide2.ps}

\plotone{finalik.eps}

\plotone{finalesp.eps}

\end{document}